\documentclass[runningheads]{paper1210}
\usepackage{graphicx}
\usepackage{multirow}
\usepackage{amsmath,graphicx,booktabs,amssymb,stmaryrd}
\usepackage{caption}
\usepackage{color}
\usepackage{bbm}
\usepackage{bbding}
\usepackage{booktabs}
\usepackage[dvipsnames]{xcolor}

\usepackage[breaklinks=true,colorlinks,bookmarks=false]{hyperref}

\begin{document}
\title{Dual-Consistency Semi-Supervised Learning with Uncertainty Quantification for COVID-19 Lesion Segmentation from CT Images}

\author{Yanwen Li*\inst{1},
Luyang Luo*\inst{2},\\
Huangjing Lin\inst{1,2},
Hao Chen\inst{3}\Envelope,
Pheng-Ann Heng\inst{2,4}}

\institute{$^1$Imsight AI Research Lab, Shenzhen, China\\
\email{liyanwen@imsightmed.com}\\
$^2$Department of Computer Science and Engineering,\\
The Chinese University of Hong Kong, Hong Kong, China\\
\email{lyluo@cse.cuhk.edu.hk}\\
$^3$Department of Computer Science and Engineering,\\
The Hong Kong University of Science and Technology, Hong Kong, China.\\
\email{jhc@cse.ust.hk}\\
$^4$Guangdong-Hong Kong-Macao Joint Laboratory of Human-Machine Intelligence-Synergy Systems, Shenzhen Institutes of Advanced Technology, Chinese Academy of Sciences, China}

\authorrunning{Yanwen Li, Luyang Luo, et al.}

\titlerunning{UDC-Net: Uncertainty-guided Dual-Consistency Learning}

\maketitle              

\footnotetext[1]{The first two authors contributed equally.}

\begin{abstract}
The novel coronavirus disease 2019 (COVID-19) characterized by atypical pneumonia has caused millions of deaths worldwide. Automatically segmenting lesions from chest Computed Tomography (CT) is a promising way to assist doctors in COVID-19 screening, treatment planning, and follow-up monitoring. However, voxel-wise annotations are extremely expert-demanding and scarce, especially when it comes to novel diseases, while an abundance of unlabeled data could be available. To tackle the challenge of limited annotations, in this paper, we propose an uncertainty-guided dual-consistency learning network (UDC-Net) for semi-supervised COVID-19 lesion segmentation from CT images. Specifically, we present a dual-consistency learning scheme that simultaneously imposes image transformation equivalence and feature perturbation invariance to effectively harness the knowledge from unlabeled data. We then quantify the segmentation uncertainty in two forms and employ them together to guide the consistency regularization for more reliable unsupervised learning. Extensive experiments showed that our proposed UDC-Net improves the fully supervised method by 6.3\% in Dice and outperforms other competitive semi-supervised approaches by significant margins, demonstrating high potential in real-world clinical practice. \footnote[2]{Code is available at \url{https://github.com/poiuohke/UDC-Net}.}

\keywords{COVID-19 \and Semi-supervised learning \and Uncertainty \and Segmentation.}
\end{abstract}

\section{Introduction}
\label{sec:introduction}

By the end of 2020, the coronavirus disease 2019 (COVID-19) \cite{zhu2020novel} characterized by atypical pneumonia has spread over 220 countries and areas, infected more than 81 million people, and caused near 1.8 million losses of lives\footnote{\url{https://covid19.who.int}}. 
For early screening of the COVID-19, chest computed tomography (CT) plays a vital role as a noninvasive and fast technique, which is reported to have high sensitivity for detecting COVID-19-related abnormal findings \cite{fang2020sensitivity,ai2020correlation,liang2020early,huang2020clinical}. To improve the screening efficiency and alleviate radiologists' reading burden, various automatic COVID-19 chest CT analysis methods have been proposed from whole-volume classification and triaging \cite{oh2020deep,wang2020covid,jin2020development,mei2020artificial,di2020hypergraph}, weakly-supervised lesion localization \cite{ma2020active,wang2020weakly}, to accurate segmentation of lesion regions \cite{fan2020inf,wang2020noise}. Among previous studies, segmentation of COVID-19 often provides more accurate descriptions of the lesions, which has significant potential in assisting doctors with the diagnosis, treatment planning, and follow-up monitoring.

Currently, advanced segmentation methods are often fully supervised and heavily rely on pixel-wise or voxel-wise annotations. For novel diseases like COVID-19, acquiring such annotations is extremely expertise-demanded and time-consuming, while unlabeled data are often abundant due to increasing positive cases.
Therefore, semi-supervised learning (SSL) that utilizes both labeled and unlabeled data is of great value to develop robust and accurate COVID-19 lesion segmentation algorithms. 
Thus far, many SSL approaches have been developed and successfully applied to various tasks \cite{van2020survey}.
Many works \cite{qiao2018deep,miyato2018virtual,berthelot2019mixmatch,ke2019dual,liu2020semi} adopts the smoothness assumption that two data samples that are close in the input space share the same label. 
This assumption is further expanded to the deep feature space, where similarities of feature maps are used for cluster assignment \cite{wang2020towards,ouali2020semi,wang2021deep}. 
Despite the achievement, these approaches do not ensure robust learning from samples with low uncertainty. 
To reduce the influence of uncertain samples, uncertainty guidance has been introduced into the literature of SSL \cite{yu2019uncertainty,xia2020uncertainty,wang2020ud,luo2020deep}.
Nevertheless, semi-supervised segmentation of COVID-19 lesions remains a challenging task, of which the annotations are extremely scarce, and the lesions often have irregular and ambiguous contours. 

To tackle the above challenges, we propose a novel deep neural network with a uncertainty-guided dual-consistency learning scheme for COVID-19 lesion segmentation from chest CT scan volumes. 
Specifically, we impose \emph{image-level transformation equivalence} out of the observation that the prediction of a sample should obtain the same transformation of the input. 
Meanwhile, we adopt \emph{feature-level perturbation invariance} to a multi-decoder V-Net, where auxiliary decoder paths take perturbated features as inputs and form output consistency with a main decoder.
Dual-consistency comprehensively enforces smoothness assumption into the SSL model from both input space and feature space, and hence the network could learn more invariant representations to diverse input or feature variants. 
Moreover, deep neural networks could memorize and easily overfit to noisy and uncertain contour points of COVID-19 lesions \cite{zhang2017understanding}, which leads to poor generalization in real-world clinical practice. 
Hence, we further introduce a novel uncertainty guidance to the consistency learning process.
Particularly, we quantify both the confidence uncertainty and the consensus uncertainty based on the multi-decoder structure.
The estimated uncertainties are then used together in an indicator function to filter out uncertain samples during training. 
The proposed uncertainty-guided dual-consistency network (UDC-Net) is evaluated on a large-scale COVID-19 dataset with 852 whole-volume chest CT scans. 
Extensive experiments show that our approach outperforms other competitive SSL-based segmentation approaches, yielding state-of-the-art performance on semi-supervised COVID-19 lesion segmentation.

\section{Method}
\label{sec:method}

\begin{figure}[t]
\centering
\includegraphics[width=\textwidth]{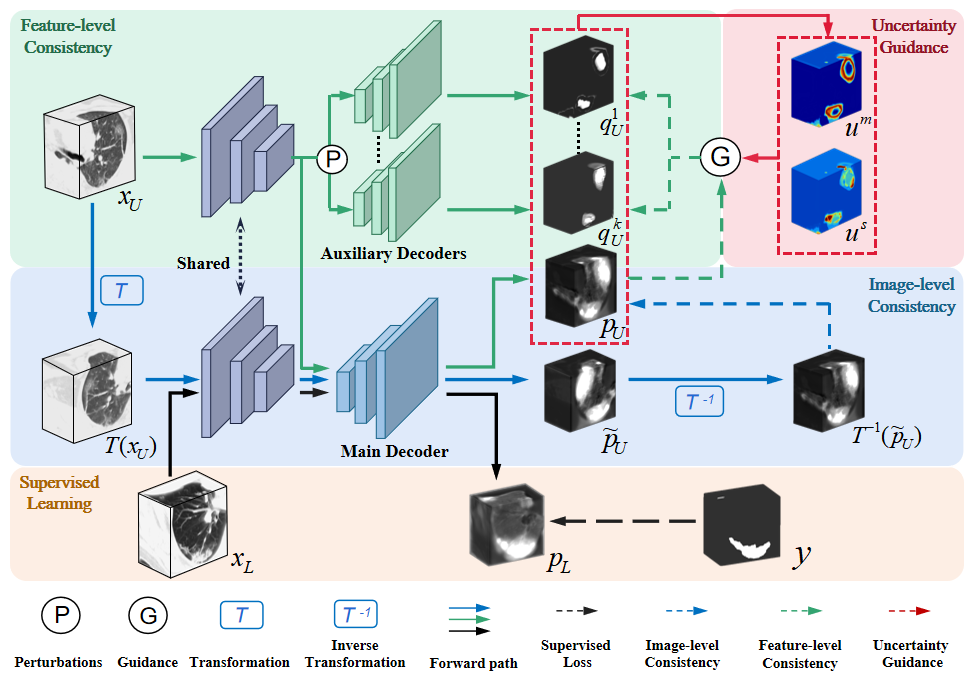}
\caption{Overview of UDC-Net. Feature-level consistency (\textcolor{green}{in green}) is formed by the main decoder's prediction $\rm p_{U}$ and auxiliary decoders' predictions \{$\rm q_{U}^{1}, \cdots, q_{U}^{k}$\}.  Image-level consistency (\textcolor{blue}{in blue}) is formed by $\rm p_{U}$ and the prediction $\rm \tilde{p}_{U}$ of transformed image. 
The confidence uncertainty $\rm u^{m}$ and the consensus uncertainty $\rm u^{s}$ are quantified by mean and standard deviation of the multi-decoders' predictions, which are then used to guide the consistency learning (\textcolor{red}{in red}).
A supervised loss is also used on the labeled data (\textcolor{Apricot}{in orange}).}
\label{fig:framework}
\end{figure}

As shown in Fig.~\ref{fig:framework}, our UDC-Net consists of a modified 3D multi-decoder V-Net \cite{milletari2016v} as its backbone. Apart from the supervised loss, our method makes full use of the unlabeled data by both feature-level and image-level consistency modules. Moreover, both the confidence uncertainty and the consensus uncertainty are estimated to guide more robust consistency learning. .

\subsection{Dual-consistency Learning for Semi-supervised Segmentation}

\textbf{Image-level Consistency Learning} via transformation equivalence of deep segmentation models $f_{\rm seg}$ indicates that
while a transformation $T(\cdot)$ is applied to an input image $x$, there should be $f_{\rm seg}(T(x)) = T(f_{\rm seg}(x))$ \cite{worrall2017harmonic}.
We conduct random transformation on the images to get the perturbated version $T(x)$ as the input to our network.
Subsequently, we have the corresponding prediction $f(T(x))$ given by the V-Net and the inverse transformation to the output $T^{-1}(f(T(x)))$, which should be consistent to the output of input data without transformation $f(x)$.
Following the notations set before, let $p=f(x)$ and $\tilde{p}=f(T(x))$ ,we introduce an image-level consistency regularization by minimizing the L2 loss between the two versions of output:
\begin{equation} \label{img_consist}
    \mathcal{L}_{\rm IC} = \frac{1}{N}\sum_{i=1}^{N}\|p_{i} - [T^{-1}(\tilde{p})]_{i}\|_{2}^{2}
\end{equation}
where $i$ and $N$ are the index and the total number of voxels, respectively.
\\\\
\textbf{Feature-level Consistency Learning} via perturbation invariance 
can also enrich the learned representation of the model \cite{miyato2018virtual}. 
Particularly, different perturbated versions of the same feature maps should maintain the same predictions. 
Following \cite{ouali2020semi}, we append several auxiliary decoders to the V-Net and inject shared encoder's outputs with various types of perturbations.
Each auxiliary decoder receives a different version of the perturbated feature map, while the main decoder receives the un-perturbated feature map. 
Denoting the prediction from the main decoder as $p$, the prediction from the $k$-th auxiliary decoder as $q^{k}$, the feature-level consistency is achieved by regularizing $p$ and each $q^{k}$ as follows:
\begin{equation}\label{feat_consist_loss}
    \mathcal{L}_{\rm FC} = \frac{1}{N\cdot K}\sum_{i=1}^{N}\sum_{k=1}^{K} \|p_{i}-q^{k}_{i}\|_{2}^{2}
\end{equation}

\noindent where $K$ is the total number of extra decoders. 
Following \cite{ouali2020semi}, seven types of feature perturbations, i.e., Feature noise, Feature dropout, Object masking, Context masking, Guided cutout, Intermediate VAT, and Random dropout, were introduced to seven auxiliary decoders, respectively. Detailed descriptions of each perturbation strategy can be found in the supplementary. All extra decoders were required to generate consistent prediction with the main decoder.

\subsection{Dual Uncertainty Quantification for Robust Learning}
The perturbation of the hidden representations during the consistency learning process could amplify the feature noises and uncertainty caused by the difficulty of accurately delineating the lesion contours of COVID-19. 
To this end, we propose to quantify both the confidence uncertainty and the consensus uncertainty of the multi-decoders, to guide more robust unsupervised learning.
\\\\
\textbf{Confidence Uncertainty} indicates whether the model generates confident predictions. 
Previous works \cite{yu2019uncertainty,luo2020deep} used the entropy of the mean prediction of multiple perturbated inputs from self-ensembling models to estimate the prediction uncertainty.
In our case, this form of uncertainty can be easily quantified using the main decoder and the $K$ auxiliary decoders as below:
\begin{equation}\label{uncertainty}
    \mu_{i} =\frac{1}{K+1}\left[\left(\sum_{k=1}^{K}q_{i}^{k}\right) + p_{i}\right]
\hspace{1em} 
\text{and}
\hspace{1em}
    u^{m}_{i}=-\mu_{i} log\mu_{i}
\end{equation}
where $i$ indicates the voxel index, $K$ is the total number of auxiliary decoders, $\mu$ is the mean prediction, and $u^{m}$ is the estimated uncertainty. The higher $u^{m}_{i}$ is, the less confidence the model is on its prediction.
\\\\
\textbf{Consensus Uncertainty} indicates whether the model generates consistent predictions over multiple runs with perturbated data \cite{Lee2020Robust,ke2019dual}.
Supposing the average prediction of a suspicious infection area is high but the outputs from different branches vary severely,
this means the area is sensitive to perturbation.
By the smoothness assumption \cite{chapelle2009semi}, the predictions for the target should be robust to perturbation, and the sensitive prediction hence highly suggests a noisy sample.
Hence, we quantify the consensus uncertainty $u^{s}$ as the standard deviation over the multi-decoders' predictions:
\begin{equation}\label{uncertainty_1}
\mathit{u^{s}_{i}}=\frac{1}{K+1}\sqrt{\left[\sum_{k=1}^{K}(q^{k}_{i}-\mu_{i} )^{2}\right]+(p_{i}-\mu_{i})^{2}}
\end{equation}

Here, $u^{s}$ essentially indicates the consensus among different decoders, which is complementary with $u^{m}$ which measures the confidence of the model.

\subsection{Uncertainty-guided Dual-consistency Learning for Segmentation}

The quantified uncertainties are used to filter out uncertain voxels and consequently guide the model to learn from more reliable unlabeled data. 
Denoting $i$ as the voxel index for the prediction volume, the reliable voxels are selected from a set $\Omega = \{i|\mathit{u^{s}_{i}}<\tau^{s} \: \& \: \mathit{u^{m}_{i}}<\tau^{m}\}$, where $\tau^{s}$ and $\tau^{m}$ are two thresholds. The cross consistency loss among decoders is then guided by:

\begin{equation}\label{uncertainty-guided-feature-consistency}
    \mathcal{L}_{\rm UFC} = \sum_{k=1}^{K} \sum_{i \in \Omega} \|p_{i}-q_{i}^{k}\|_{2}^{2}
\end{equation}

Here, the uncertainty guidance is applied onto feature-level consistency learning as the uncertainties are generated with feature perturbations. Thus, the total loss for our uncertainty-guided dual-consistency learning UDC-Net for semi-supervised lesion segmentation is as follows:

\begin{equation}\label{loss_supervise}
    \mathcal{L} = \mathcal{L}_{\rm S} + \alpha\mathcal{L}_{\rm IC } + \beta\mathcal{L}_{\rm UFC}
\end{equation}

\noindent where $\mathcal{L}_{\rm S}$ is the supervised loss consists of a Dice loss and a cross entropy loss, $\alpha$ and $\beta$ are two hyper-parameters weighing the contributions from different losses.

During training, we first trained a supervised V-Net and then added the extra decoders for finetuning with uncertainty-guided consistency learning.
The training process was terminated if the Dice coefficient on the validation dataset stagnated.
Adam \cite{kingma2014adam} was used as the optimizer with an initial learning rate of 0.001 and a learning decay rate of 0.95 per epoch. 
As widely adopted by SSL works \cite{tarvainen2017mean,ouali2020semi}, $\alpha$ and $\beta$ were set to be two sigmoid-shape monotonically functions of the training steps with maximum of 1. 
The threshold $\tau^{m}$ and $\tau^{s}$ were set to 0.34 and 0.12 after tuning on the validation set. 
For testing, we carried out sliding window inference and took only the main decoder's prediction. 
All implementation was done with Pytorch \cite{paszke2019pytorch} on an NVIDIA TITAN X GPU.

\section{Experiments}
\label{sec:experiments}

\subsection{Datasets and Evaluation Metrics}

\textbf{Datasets.} In total, 852 chest CT volumes acquired from December 2019 to April 2020 were collected and enrolled in this study, among which 144 were voxel-annotated by four experienced radiologists. 
The labeled data were divided into: (1) 65 cases as labeled training dataset; (2) 9 cases as the validation set; and (3) 70 cases as the testing set. The remained 708 chest CT scans were used as the unlabeled training data.
\\

\noindent\textbf{Evaluation Metrics.} We adopted Dice Score (DSC), Jaccard similarity cofficient (Jaccard), and Average Symmetric Surface Distance (ASD) to evaluate the segmentation performance.

\begin{table}[t!]
\begin{center}
\caption{Ablation study of different components. All results are reported as validation/testing results. (FC: feature-level consistency; IC: image-level consistency; UM: confidence uncertainty computed by the mean of the multi-decoders' predictions; US: consensus uncertainty computed by the standard deviation of the multi-decoders' predictions)}
\setlength{\tabcolsep}{3mm}{
\begin{tabular}{c|c|c|c|c|c|c}
\toprule[1pt]
\multicolumn{4}{c|}{\textbf{Components}}                                               & \multicolumn{3}{c}{\textbf{Evaluation Metrics}}                       \\ \hline
\textbf{IC} & \textbf{\begin{tabular}[c]{@{}c@{}}FC\end{tabular}} & \textbf{\begin{tabular}[c]{@{}c@{}}UM\end{tabular}} & \textbf{\begin{tabular}[c]{@{}c@{}}US\end{tabular}} & \textbf{DSC{[}\%{]} $\uparrow$} & \textbf{Jaccard{[}\%{]} $\uparrow$} & \textbf{ASD{[}mm{]} $\downarrow$} \\ \hline
             &                &                         &                           & 70.0 / 71.1               & 56.5 / 56.8                    & 12.1 / 12.1                  \\
\checkmark   &                 &                         &                           & 70.3 / 73.5              & 56.7 / 59.3                   & 12.2 / 8.4                  \\
            &  \checkmark       &                         &                           & 71.4 / 75.6               & 58.4 / 62.3                   & 12.1 / 6.1                  \\
\checkmark   & \checkmark     &                         &                           & 71.9 / 76.7                & 58.9 / 63.7                   & 11.7 / 5.8                    \\
\checkmark   & \checkmark     & \checkmark              &                           & 72.2 / 77.0               & 59.0 / 64.0                   &  11.3 / \textbf{3.2}          \\
\checkmark   & \checkmark     &               & \checkmark                          & 72.4 / 77.2               & 59.5 / 64.3                   &  11.4 / 4.1          \\
\checkmark   & \checkmark     & \checkmark              & \checkmark                & \textbf{72.7} / \textbf{77.4}      &  \textbf{59.9} / \textbf{64.5}          & \textbf{10.9} / 3.9                   \\ 

\bottomrule[1pt]
\end{tabular}}
\label{table:result_our_method_part}
\end{center}
\end{table}

\subsection{Ablation Study on Different Components}
We conduct ablation studies to analyze the contributions of our proposed methods, and the quantitative results can be seen in Table~\ref{table:result_our_method_part}.
Regarding the testing set performance, 
image-level consistency (IC) shows increases of 2.4\% in DSC, 2.5\% in Jaccard, and 3.7 in ASD comparing to 3D V-Net. 
Meanwhile, feature-level consistency (FC) regularization leads to a large improvement of 4.5\% in DSC, 5.5\% in Jaccard, and 6.0 in ASD comparing to 3D V-Net. 
Unifying dual consistencies further improves DSC and Jaccard with about 1\%, which demonstrates the effectiveness of learning from the unlabeled data.
Further, introducing either the confidence uncertainty or the consensus uncertainty guidance consistently benefit the learning of the unlabeled data. 
Moreover, our method with dual uncertainty achieves better DSC and Jaccard with a comparable ASD to those of the single-uncertainty models, further demonstrating that dual uncertainties are complementary for guiding more robust learning.

\begin{table}[b!]
\centering
\caption{Quantitative comparison with other semi-supervised methods.}
\begin{tabular}{c|ccc}
\hline
\toprule[1pt]
\multirow{3}{*}{Methods} & \multicolumn{3}{c}{\textbf{Evaluation Metrics}}                                                                                                          \\ \cline{2-4} 
                         & \multicolumn{1}{c|}{\textbf{DSC{[}\%{]} $\uparrow$}}                      & \multicolumn{1}{c|}{\textbf{Jaccard{[}\%{]} $\uparrow$} }                  & \multicolumn{1}{c}{\textbf{ASD{[}mm{]} $\downarrow$}} \\ \cline{2-4}  \hline
V-Net \cite{milletari2016v}                    & \multicolumn{1}{c|}{71.1 $\pm$ 0.40} & \multicolumn{1}{c|}{56.8 $\pm$ 0.45} &  12.1 $\pm$ 2.1 \\
Mean Teacher~\cite{tarvainen2017mean}             &  \multicolumn{1}{c|}{72.5 $\pm$ 0.25} & \multicolumn{1}{c|}{58.2 $\pm$ 0.36} & \multicolumn{1}{c}{11.3 $\pm$ 1.8}\\
UA-MT~\cite{yu2019uncertainty}                    & \multicolumn{1}{c|}{74.0 $\pm$ 0.11}  & \multicolumn{1}{c|}{60.1 $\pm$ 0.15} & \multicolumn{1}{c}{9.2 $\pm$ 0.9} \\
TCSM~\cite{li2020transformation}                     & \multicolumn{1}{c|}{72.9 $\pm$ 0.46} & \multicolumn{1}{c|}{58.9 $\pm$ 0.58} & \multicolumn{1}{c}{9.1 $\pm$ 1.4}\\
CCT~\cite{ouali2020semi}                       & \multicolumn{1}{c|}{75.6 $\pm$ 0.11} & \multicolumn{1}{c|}{62.3 $\pm$ 0.19} & \multicolumn{1}{c}{6.1 $\pm$ 0.7}  \\
UDC-Net(ours)            & \multicolumn{1}{c|}{\textbf{77.4 $\pm$ 0.14}} & \multicolumn{1}{c|}{\textbf{64.5 $\pm$ 0.15}} & \multicolumn{1}{c}{\textbf{3.9 $\pm$ 0.5}} \\ 

\bottomrule[1pt]
\end{tabular}
\label{table:result_compared_with_others}
\end{table}

\subsection{Comparison with State-of-the-art Methods}
\begin{figure}
\centering
\makebox[\textwidth][c]{\includegraphics[width = \linewidth]{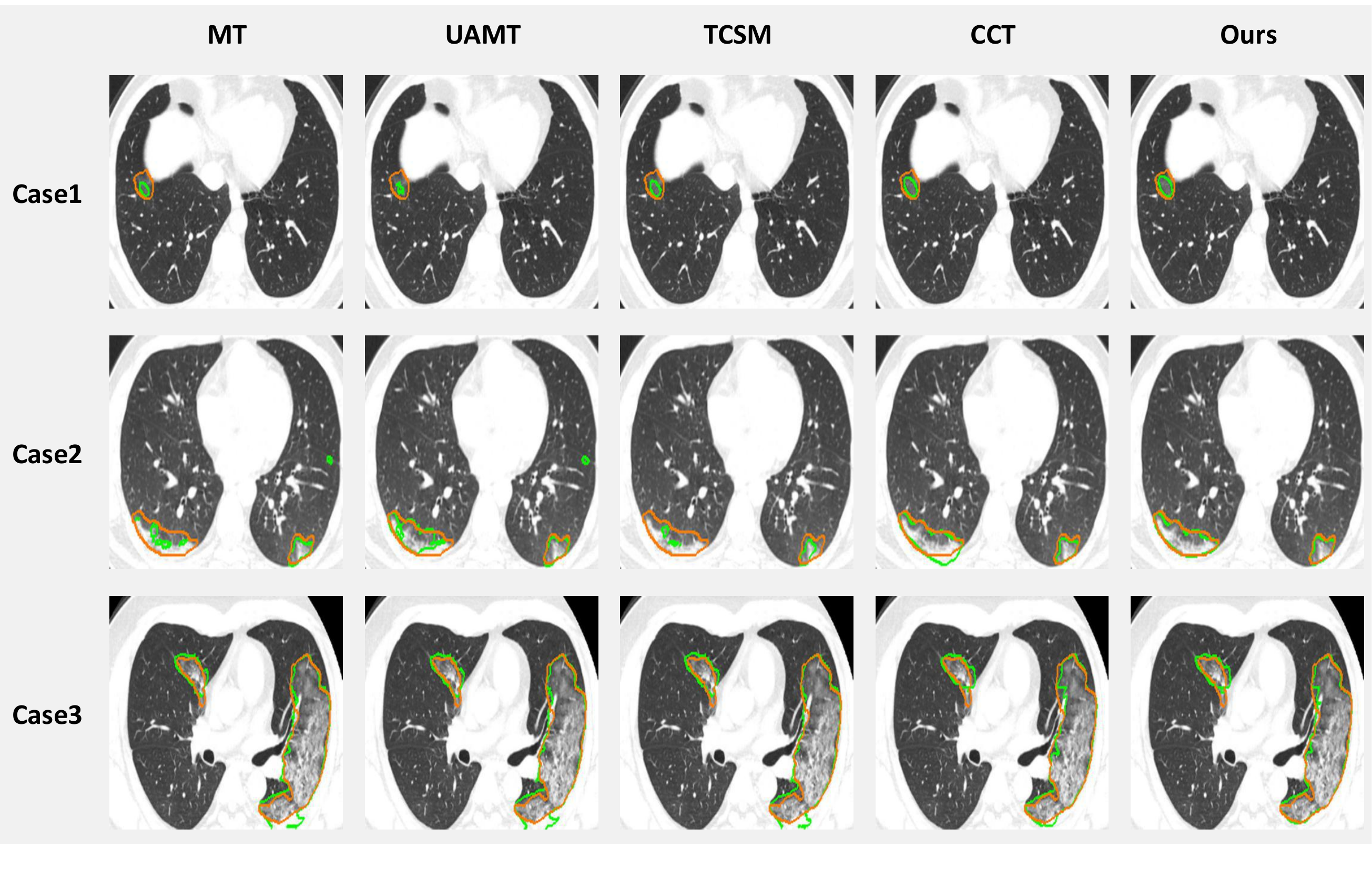}}
\centering
\caption{Qualitative comparison. \textcolor{green}{Green} and \textcolor{Orange}{orange} curves delineate the model prediction and ground truth, respectively. Best viewed in color.}
\label{2D_qualitative}
\end{figure}

We compare our method against other state-of-the-art semi-supervised segmentation approaches. Several recent models were implemented, including Mean-Teacher (MT) \cite{tarvainen2017mean}, Uncertainty-aware mean teacher \cite{yu2019uncertainty}, Transformation-consistent Self-ensembling Model (TCSM) \cite{li2020transformation}, and Cross Consistency Training (CCT) \cite{ouali2020semi}. We run each methods four times with different random seeds.\\\\
\textbf{Quantitative comparison} results are reported in Table~\ref{table:result_compared_with_others}. For a fair comparison, we implemented all methods with the 3D V-Net as backbone.
As observed, UDC-Net outperforms all other methods with at least 1.8\% in Dice, 2.2\% in Jaccard, and 2.2 in ASD, showing outstanding unsupervised learning efficacy.\\\\
\textbf{Qualitative comparison} is illustrated by visualizing the segmentation results in Figure~\ref{2D_qualitative}. As demonstrated, Our UDC-Net delineates more accurate lesion contours than other methods regarding diverse shapes and sizes of lesion.
Visualization of the two uncertainties can be found in the supplementary.

\begin{table}[b!]
\caption{Quantitative performance comparison under different numbers of training labeled/unlabeled data.}
\centering
\begin{tabular}{c|c|c|c|c|c}
\toprule[1pt]
\multirow{2}{*}{\textbf{Method}} & \multicolumn{2}{c|}{\textbf{\# scans used}} & \multicolumn{3}{c}{\textbf{Evaluation Metrics}}                                   \\ \cline{2-6} 
                                 & \textbf{Labeled}    & \textbf{Unlabeled}    & \textbf{DSC{[}\%{]} $\uparrow$} & \textbf{Jaccard{[}\%{]} $\uparrow$} & \textbf{ASD{[}mm{]} $\downarrow$} \\ \hline
V-Net~\cite{milletari2016v}                            & 32                & 0                     & 70.4                 & 56.0                       & 4.3                     \\ \hline
CCT~\cite{ouali2020semi}                             & 32                & 140                   & 74.4                 & 60.7                     & 5.8                     \\
\textbf{UDC-Net (ours)}                             & 32                & 140                   & \textbf{75.0}                   & \textbf{61.5}                     & \textbf{4.8}                     \\ \hline
CCT~\cite{ouali2020semi}                                & 32                & 708                   & 75.2                 & 61.6                     & 7.0                     \\
\textbf{UDC-Net (ours)}                               & 32                & 708                   & \textbf{76.9}                 & \textbf{64.0}                     & \textbf{4.4}                     \\ \hline \hline
V-Net                            & 65                & 0                     & 71.1                 & 56.8                     & 12.1                    \\ \hline
CCT~\cite{ouali2020semi}                                & 65                & 140                   & 75.1                 & 61.8                     & 5.8                     \\
\textbf{UDC-Net (ours)}                               & 65                & 140                   & \textbf{76.6}                 & \textbf{63.5}                     & \textbf{5.4}                     \\ \hline
CCT~\cite{ouali2020semi}                                & 65                & 708                   & 75.6                 & 62.3                     & 6.0                     \\
\textbf{UDC-Net (ours)}                               & 65                & 708                   & \textbf{77.4}                 & \textbf{64.5}                     & \textbf{3.9}                     \\
\bottomrule[1pt]
\end{tabular}
\label{table:result_different_ratio}
\end{table}

\subsection{Analysis on Efficacy of Leveraging Unlabeled Data}
We further evaluate our UDC-Net's effectiveness by varying the ratios of labeled and unlabeled training data.
Table~\ref{table:result_different_ratio} shows that UDC-Net consistently improves the baseline V-Net with significant margins in both DSC, Jaccard, and ASD, whenever 32 or 65 labeled scans are provided. Moreover, the proposed approach consistently outperforms CCT \cite{ouali2020semi} (the best model among those compared with ours) under all different scenarios. Notably, when less data are given, UDC-Net shows comparable or even better results than CCT. For instance, UDC-net achieves 75.0\% DSC, 61.5\% Jaccard, and 4.8 ASD with 32 labeled scans and 140 unlabeled scans (3rd row), which is comparable to the performance of CCT with double labeled scans (7th  row). With 65 labeled scans and 140 unlabeled scans, UDC-Net (8th row) shows superior performance than CCT with 5 times unlabeled data (9th row). These findings demonstrate that our method enables more efficient unsupervised learning, suggesting 

\section{Conclusions}
\label{sec:conclusion}

In this paper, we present an uncertainty-guided dual-consistency learning method for semi-supervised COVID-19 lesion segmentation from chest CT scans. Image-level transformation equivalence and feature-level perturbation invariance are both introduced to form dual consistency learning from unlabeled data. Meanwhile, the dual uncertainty mechanism further improves the learning process with more reliable and robust guidance. 
Extensive experiments on a large COVID-19 dataset demonstrate the efficiency of our method in real-world scenarios. 
Future work will include improving the method with more robust knowledge distillation and generalizing to other semi-supervised learning tasks.

\subsubsection{Acknowledgement.} This work was supported by Key-Area Research and Development Program of Guangdong Province, China (2020B010165004), Hong Kong Innovation and Technology Fund (Project No. ITS/311/18FP and Project No. ITS/426/17FP.), and National Natural Science Foundation of China with Project No. U1813204.

%
%
%
\bibliographystyle{paper1210}
\bibliography{paper1210}

\newpage

\section{Supplementary Materials}

\begin{table}[]
\caption{List of perturbations used in  feature-level consistency learning.}
\centering
\begin{tabular}{ll}
\hline
\textbf{Perturbations}\cite{ouali2020semi} & \textbf{Description}                                                                                                                                                                                                                               \\ \hline
Feature Noise          & A noise tensor N is applied to the output of encoder z to get $\tilde{z} = z*N+z$.                                                                                                                                                                                    \\
Feature Dropout        & \begin{tabular}[c]{@{}l@{}}Generating a randomly dropout mask $M_{drop}$ to obtain perturbated\\ $\tilde{z}=z*M_{drop}$.\end{tabular}                                                                                                                                                                                   \\
Object Masking         & \begin{tabular}[c]{@{}l@{}}Generating a object mask $M_{obj}$ using the output of main decoder \\to get $\tilde{z}=z*M_{obj}$.\end{tabular}                                                                                                           \\
Context Masking        & \begin{tabular}[c]{@{}l@{}}Generating a context mask $M_{con}=1-M_{obj}$ to obtain $\tilde{z}=z*M_{con}$.\end{tabular}                                                                                                                              \\
Guided cutout          & \begin{tabular}[c]{@{}l@{}}Zero-out a random crop within each object's bounding box from \\ the corresponding feature map z.\end{tabular}                                                                                                          \\
Intermediate VAT       & \begin{tabular}[c]{@{}l@{}}Using VAT to push the distribution to be isotropically smooth. Finding \\the adversarial perturbation $r_{adv}$ alter its prediction the most and \\injected into z to obtain $\tilde{z}=r_{adv}+z$.\cite{miyato2018virtual}\end{tabular} \\
Random dropout         & Spacial dropout applied to z as a random perturbation                                                                                                                                                                                              \\ \hline
\end{tabular}
\label{table:perturbations}
\end{table}

\begin{figure}[]
\centering
\includegraphics[width = \linewidth]{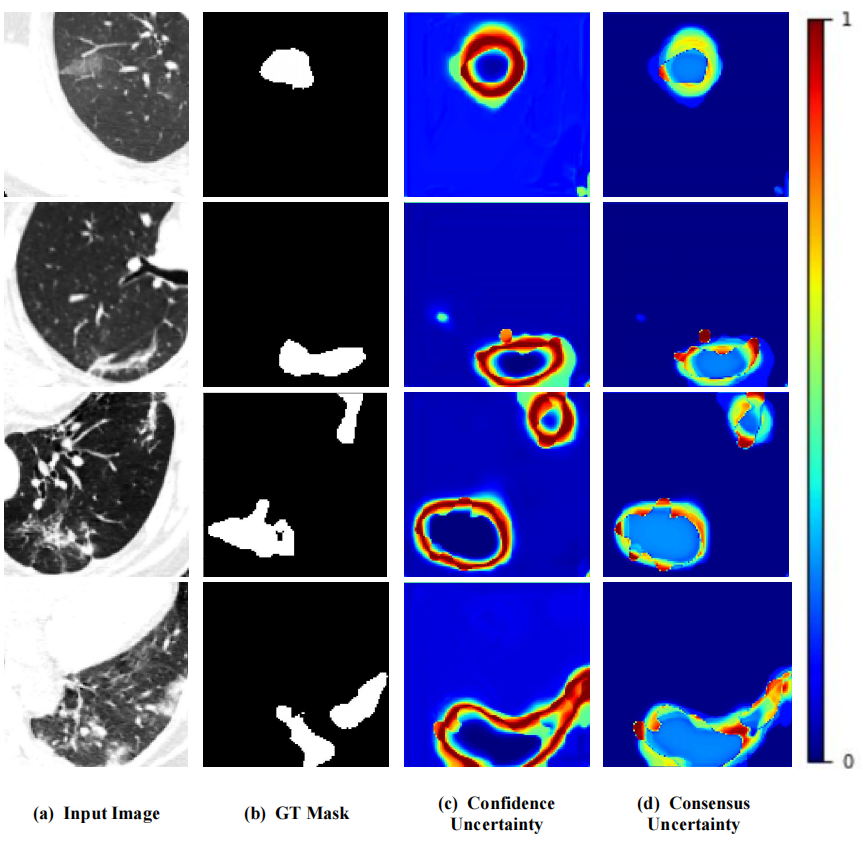}
\centering
\caption{Visualization of input images, ground truth masks, confidence uncertainty map, and consensus uncertainty map. The visualization results demonstrates that our proposed confidence and consensus uncertainties are complementary.}
\label{uncertain_maps}
\end{figure}

\begin{figure}
\centering
\includegraphics[width = 1.0\linewidth]{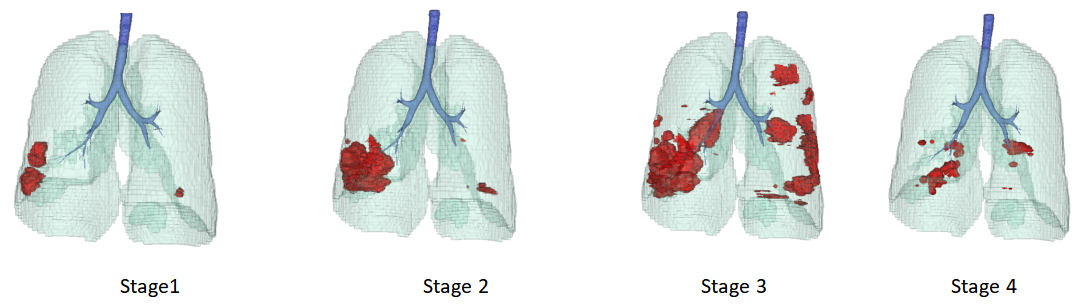}
\centering
\caption{An example of monitoring the lesion development of a patient from mild infection (stage 1), to common infection (stage 2), severe infection (stage 3), and finally to recovery stage (stage 4).}
\label{3d_view}
\end{figure}

\end{document}